# Energy efficient prediction clustering algorithm for multilevel heterogeneous wireless sensor networks

Tang Liu[a,b,c,*] Jian Peng[b,c] Jin Yang[d] Chunli Wang[e]

[a]*College of Fundamental Education, Sichuan Normal University, Chengdu 610068, China*

[b]*State key Laboratory of Networking and Switching Technology, Beijing University of Posts and Telecommunications, Beijing 100876, China*

[c]*College of Computer Science, Sichuan University, Chengdu 610065, China*

[d]*Department of Computer Science, Leshan Normal University. Leshan 614000, China*

[e] *English Department, Chengdu University of Information Technology, Chengdu 610225, China*

**Abstract:** In designing wireless sensor networks, it is important to reduce energy dissipation and prolong network lifetime. In this paper, a new model with energy and monitored objects heterogeneity is proposed for heterogeneous wireless sensor networks. We put forward an energy-efficient prediction clustering algorithm, which is adaptive to the heterogeneous model. This algorithm enables the nodes to select the cluster head according to factors such as energy and communication cost, thus the nodes with higher residual energy have higher probability to become a cluster head than those with lower residual energy, so that the network energy can be dissipated uniformly. In order to reduce energy consumption when broadcasting in clustering phase and prolong network lifetime, an energy consumption prediction model is established for regular data acquisition nodes. Simulation results show that compared with current clustering algorithms, this algorithm can achieve longer sensor network lifetime, higher energy efficiency and superior network monitoring quality.

## 1 Introduction

Over recent years, wireless sensor networks (WSN) [1] with nodes equipped with a large number of small energy devices have become a hot research and have a wide range of potential applications including environmental monitoring, military detection, health monitoring, industrial control and home networks [2-5]. But in practical applications, in order to meet the demands of various applications for the technologies of sensor networks, increasing attentions have been attracted to the researches on heterogeneous wireless sensor networks HWSN [6].

HWSN is composed of different types of sensor nodes, which are in a wide range of applications [7,8]. In fact, the heterogeneity is common in the wireless sensor networks [9]. For HWSN, it should be given priority to reduce energy dissipation in network operation, improve network load and stability and prolong the network lifetime.

Energy consumption in networks can be effectively reduced by organizing clustering sensor nodes, so many energy-efficient routing protocols are designed on the basis of the clustering structure. Currently, a number of distributed clustering protocols are proposed. In accordance with the networks, homogeneous or heterogeneous, to which the protocols are adaptive, clustering protocols can be categorized into homogeneous clustering protocols and heterogeneous clustering

* Corresponding author. College of Fundamental Education, Sichuan Normal University, No. 5, North JingAn Road, Chengdu 610068, PR China. Tel./fax: +86 13880703601.

*E-mail addresses:* crikey@163.com (T. Liu), penguest@163.com (J.Peng), jinnyang@163.com (J.Yang), lizzywang@cuit.edu.cn (C. Wang).

protocols. Due to the dynamic and complex nature of energy configuration and network evolution, it is very difficult to design a clustering protocol which can save energy and provide reliable data transmission in heterogeneous networks.

Environmental monitoring applications usually involve a variety of data collection within the same monitoring range. Some data collection is of regularity. For instance, informations such as temperature and humidity should be reported at a fixed interval, and the data sent every time is of equal length, while the report of fire data does not show such regularity and each time, the length of reported data is different according to the degree of the fire.

In this paper, a new heterogeneous sensor networks model with heterogeneity of monitored objects and energy heterogeneity of all nodes is proposed. For the heterogeneous networks with such properties, in order to make more rational use of network energy and prolong the lifetime of the networks, this paper presents an Energy-Efficient Prediction Clustering Algorithm (EEPCA).

EEPCA gets informed of the mutual distance between nodes through broadcasting in the initial stage of nodes clustering. It determines node energy factor by comparing the energy of a node with the average energy of other nodes within the communication range and determines communication cost factor according to the ratio of the average energy consumed in one communication within all nodes and the ideal average energy consumption after the node becomes the cluster head. The probability for nodes to become cluster heads is directly related to energy factor and communication cost factor. All nodes in the networks take turns as cluster heads to achieve uniform energy consumption. In order to save energy consumed by broadcasting energy information in each round of nodes clustering, an energy predication model is established for nodes whose data collection (such as temperature, humidity, etc) is of regularity in time interval and message length. Considering the changes in networks environment and errors between calculated and actual node energy consumption, set the nodes do not need to broadcast their energy information if the difference between the node residual energy in the initial stage at the current round and the predicted value at the last round is within a certain range. Simulation results show, EEPCA can achieve longer lifetime, higher energy efficiency and superior network monitoring quality compared with other clustering protocols such as LEACH, SEP and EDFCM.

The rest of this paper is organized as follows. In Section 2, the related work is discussed. In Section 3, we present the networks model, energy model and the performance measures for wireless sensor networks. Section 4 exhibits the details of EEPCA. In Section 5, we evaluate the performance of EEPCA via simulations and compare the results with LEACH,SEP and EDFCM. Finally, Section 6 concludes the paper and future work is pointed out.

## 2 Related Work

LEACH [10] is one of the most popular distributed cluster-based routing protocols in wireless sensor networks. LEACH assumes that all nodes are schemed with the same initial energy and each node $i$ generates a random probability between 0-1. Network operation time is divided into many time slots, known as round. One round consists of two stages, namely initialization and stability. In the initialization phase, LEACH carries out cluster head selection. In order to balance the load of all network nodes, LEACH elects about cluster head nodes in each round, where is the proportion of optimal cluster heads. If the probability of a node $i$ is less than the following probability threshold, it becomes the cluster head.

$$T(i)=\begin{cases}\dfrac{p_{opt}}{1-p_{opt}\left(r \bmod \dfrac{1}{p_{opt}}\right)}, & \text{if } i\in G\\ 0, & \text{otherwise}\end{cases} \tag{1}$$

Where, $r$ is the current number of rounds, $G$ is a set of cluster head nodes which fail to make cluster heads in the latest round $\left(r \bmod \left(1/p_{opt}\right)\right)$.

When cluster heads are chosen, all these head nodes broadcast this message to other nodes. According to the strength of received messages, nodes determine which cluster head they would join and inform the corresponding cluster head. Based on TDMA approach, cluster heads allocate time slot to cluster members and the networks proceeds into a stable phase, in which each node sends the monitored data back to the cluster head node in the corresponding slot and the cluster head node transfers the received data to the Base Station (BS) after aggregation. So far, one round comes to the end and starts the next. In this way, each node has opportunity to become a cluster head node dissipating more energy.

However, LEACH has some constraints, including: (1) it does not take into account the optimization of the number of cluster heads. The probability for a random node to become a cluster head is $p$, and therefore the number of cluster heads is proportional to the number of nodes; (2) as cluster heads are randomly selected, and therefore LEACH can not guarantee cluster heads are uniformly distributed in the networks. Meanwhile, the probability threshold does not take into account the energy factor. LEACH algorithm therefore must base itself on two assumptions so as to achieve uniform energy consumption at per node: (1) the initial energy of each node is equal; (2) the energy consumed at each node when acting as the cluster head is equal. Therefore, it is difficult to apply LEACH algorithm to an actual networks application.

Many researchers have done profound work probing into HWSN.

In [10], authors improved LEACH algorithm and put forward an algorithm of electing cluster heads according to the residual energy. However, each node needs to know the total energy of the current network to determine whether it can become the cluster head, which requires support of routing protocols and therefore distributed implementation is difficult to achieve. This algorithm is called LEACH-C. SEP [11] is designed for two-level heterogeneous networks in which there are merely two kinds of nodes with different initial energies. But in the multi-level heterogeneous networks, nodes' initial energy is randomly determined within a certain range, so SEP does not suit for such a heterogeneous environment. For further researches, a heterogeneous network model in term of different initial energies is discussed in [6,12-14]. In [15], the authors introduced a cluster head election method using fuzzy logic to overcome the defects of LEACH. They investigated that the network lifetime can be prolonged by using fuzzy variables.

In [16], authors proposed EEHC protocol. This protocol selects cluster heads based on the weighted probability of each node related to the initial energy, the more initial energy, the higher probability the node will be selected as a cluster head. However, this protocol can not predict energy consumption, so its performance is limited in heterogeneous networks in which part of nodes are regular data acquisition nodes.

In [17], authors proposed EDFCM protocol, which applies to networks with three different kinds of heterogeneous nodes. Nodes in the networks model of this protocol fall into two ordinary types: one performing the function of managing information and the other collecting different data(type_0, type_1). type_1 have more complex hardware and software architectures, so it has more initial energy and greater data transfer capability. To guarantee an optimum number of cluster heads selected in actual operations, authors propose a stable selection and reliable transmission protocol based on a method of energy dissipation forecast and clustering management. But the application of this protocol is limited to the networks with only two types of ordinary nodes.

In [18], authors proposed ERP clustering routing protocol for HWSN. In this paper, an evolutionary algorithm with an appropriate fitness function is proposed with the intrinsic properties of clustering in mind. Main idea of the proposed ERP is the incorporation of compactness and separation error criteria in the fitness function to direct the search into promising solutions. Against LEACH and HCR, ERP can prolong network lifetime and stability period. However, compared with SEP, ERP gains longer network lifetime, but at the expense of less stability awareness.

**3 System Model and Problem Description**

3.1 Heterogeneous model for wireless sensor networks

To meet the demands of efficient environmental monitoring, we describe our HWSN model with both different initial energies and monitored objects. The basic assumptions of networks model: the networks is located in a M × M square area(Fig. 1), N sensor nodes are randomly distributed within the networks, nodes are slightly mobile or stationary, and base station is located in the middle of the area. The networks perform the task of environmental monitoring and sensor nodes monitor a variety of objects. Define nodes monitoring temperature, humidity, wind direction etc. as regular data acquisition (RDA) nodes; these nodes send back messages of fixed length at a fixed interval; nodes monitoring fire are not regular in acquiring data and the messages sent back are not regular.

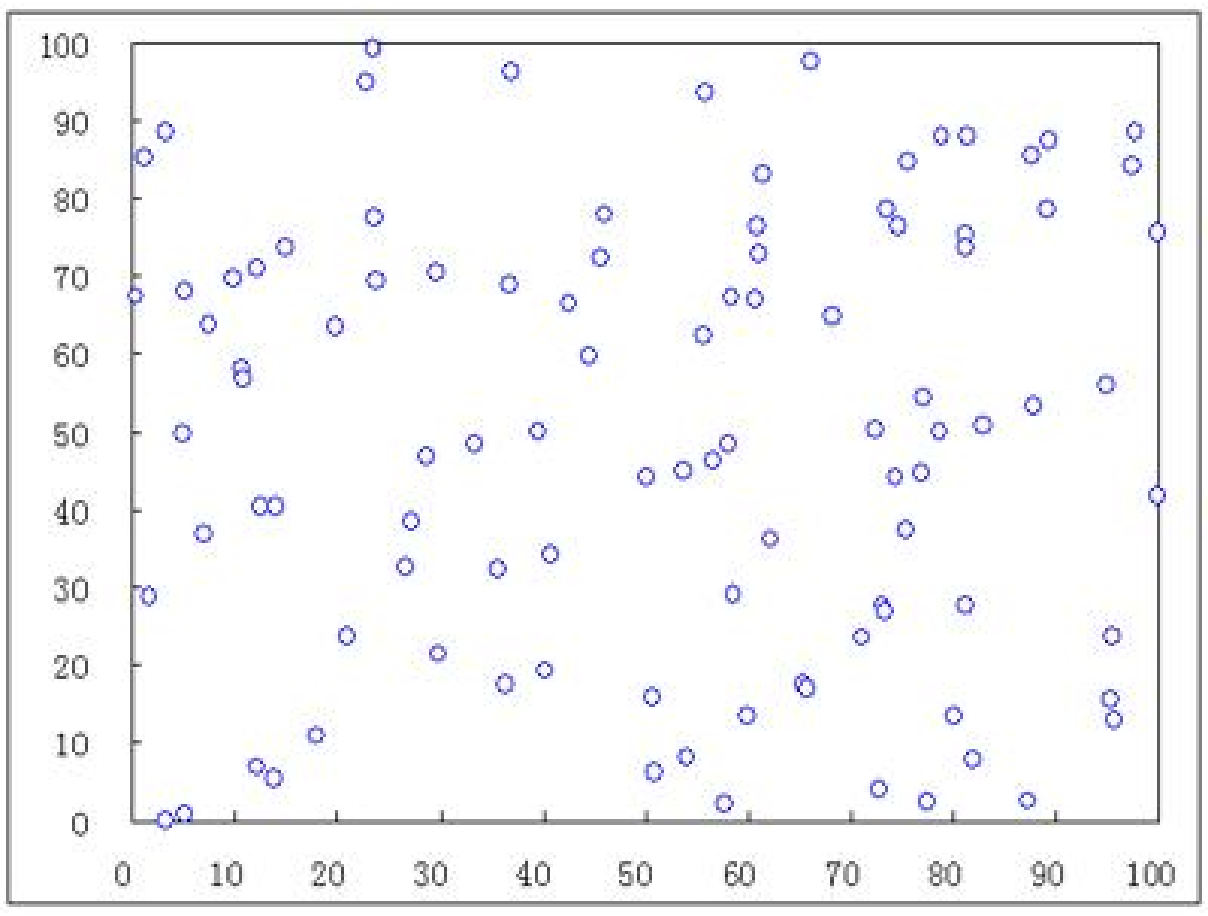

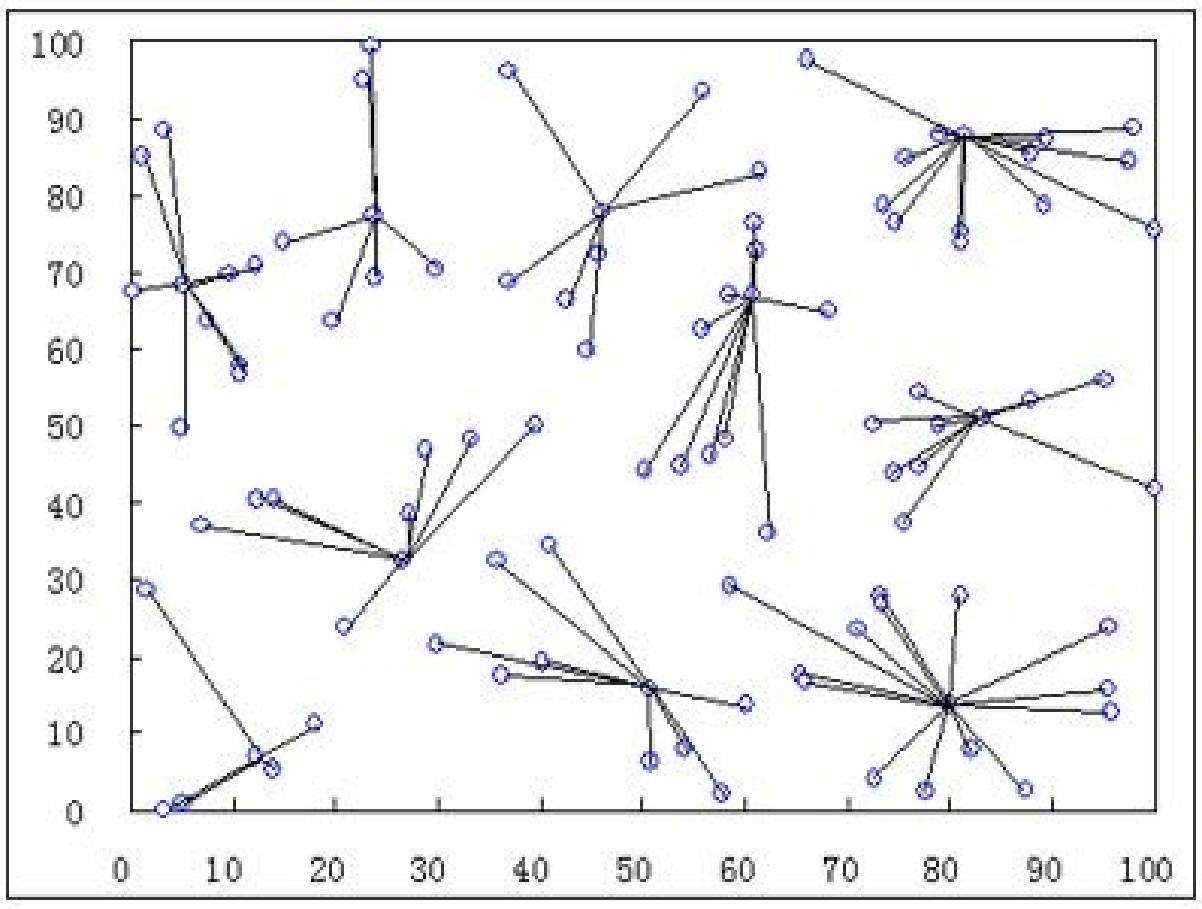

Fig. 1. (a) 100-node random heterogeneous network. (b) Dynamic clustering structure by EDFCM.

Therefore, nodes are heterogeneous in two ways: (1) heterogeneous data-acquisition -regularity: some nodes are regular in acquiring data and some are not. All regular nodes send $n_1 \sim n_2$ times messages in a rotation cycle times and the message sizes are between $[l_1, l_2]$ bits; (2) the initial energy of all nodes are heterogeneous.

Nodes communication links are symmetric and nodes do not have any location information, but they can calculate the distance between nodes according to signal strength received. Nodes in the networks are organized in the form of clusters. Cluster heads perform the function of data fusion and are responsible for the resultant data transmission to the BS. There is only one BS in the networks and wireless transmission power is controllable.

Node initial energy is randomly distributed in the closed interval $[E_{\min}, E_{\max}]$, where $E_{\min}$ is the lower bound of the energy, $E_{\max}$ determines the value of maximum node initial energy. For any node $i$, its initial energy is $E_i$.

3.2 Energy Models

This article applies a simple energy consumption model [10] to calculate energy consumption in communication, ignoring energy consumption of nodes in the process of computing, storage, etc. In the process of transmitting $l$ bits message through distance $d$, the energy consumption of the transmitter is:

$$E_{Tx}(l,d) = E_{Tx_elec}(k) + ET_{Tx_amp}(l,d) = \begin{cases} lE_{elec} + l\varepsilon_{fs}d^2, d < d_0 \\ lE_{elec} + l\varepsilon_{mp}d^4, d \geq d_0 \end{cases} \tag{2}$$

Receiver's energy consumption is

$$E_{Rx}(l) = E_{Rx_elec}(l) = lE_{elec} \tag{3}$$

Where $E_{elec}$ is the energy dissipated per bit to run the transmitter or the receiver circuit, and $\varepsilon_{fs}d^2$ and $\varepsilon_{mp}d^4$ are the amplifier energy that depend on the transmitter amplifier model.

3.3 Problem Description

Essentially, all WSN clustering algorithms are intended to solve the problem of unbalanced networks load, and to achieve uniform distribution of energy dissipation at all nodes, so as to prolong the network lifetime as much as possible. Therefore, EEPCA must take full account of the following:

(1) algorithm should be fully distributive and self-organized. Nodes determine their own state based only on local information, and each node must decide whether to become a cluster head or a member belonging to a cluster in the clustering phase [10];

(2) nodes with more residual energy must have higher probability to become cluster head and it must be ensured that the cluster has a smaller communication cost, but energy is not the only factor for cluster head selection;

(3) cluster load balancing must be ensured;

(4) EEPCA operates in rounds. In order to save energy consumption when nodes broadcast in initial clustering phase of each round, an energy prediction model of RDA nodes is established.

## 4. EEPCA Clustering Algorithm

4.1 Calculation of distance between Nodes

Nodes in the networks can perceive their mutual distance according to attenuation of signal strength in the process of transmission. In clustering phase, all nodes use certain transmission energy for broadcast. For instance, with energy $E_i^{tran}$, node $i$ broadcasts information to other nodes, including its message sending cycle $t_i$, message length $l_i$ and its energy information $E_i$. Node $j$ detects the received signal strength (received energy) $E_{j,i}^{rec}$ while receiving messages. The relationship between transmission energy and reception energy is as follows [20]:

$$E_{j,i}^{rec} = \frac{K}{d_{i,j}^{\alpha}} \times E_i^{tran} \tag{4}$$

Where $K$ is a constant, $d_{i,j}^{\alpha}$ is the relative distance between node $i$ and node $j$. $\alpha$ is distance - energy gradient, and its value varies from 1 to 6 according to the physical environment in which the sensor networks operate. Thus, the distance between $i$ and $j$ is:

$$d_{i,j} = \sqrt[\alpha]{\frac{K \times E_i^{tran}}{E_{j,i}^{rec}}} \tag{5}$$

The node establish a routing table of neighboring nodes based on received data and save all relevant information of all nodes within its communication range. All nodes in the networks are marked by the only integer value, which is each node's ID. The information stored in the routing table includes the distance between the node and its neighboring nodes, cluster head node's ID, the distance to the cluster head, the current energy and predicted energy consumption.

4.2 Cluster head selection

The cluster head node has to perform extra functions such as data fusion and relaying messages, so its energy consumption rate is much higher than that of ordinary nodes. In order to prevent some nodes from dying too soon due to excessive energy cost, the nodes with more residual energy should be given greater opportunity to become cluster heads and all nodes take their turns to be cluster head nodes.

Set $p_{opt}$ is the proportion of optimal cluster heads and $p_i$ is the probability for node i to be selected as the cluster head. Obviously, if the current energy at all nodes is equal to each other, $p_{opt} = p_i$ can ensure that all nodes die at the same time. In energy-heterogeneous WSN, $p_i$ calculation is much more complicated. Currently, many clustering algorithms in HWSN determine $p_i$ by using the ratio of nodes' current residual energy and the average energy of the entire networks, but the latter is very difficult to obtain [13], especially for networks in which different nodes are monitoring different objects. Consequently, major error is likely to happen to the estimated average energy.

Ideally, nodes are distributed uniformly and send back data at identical frequency and length. Set $d_{toBS}$ is the average distance between the head node and the BS and $d_{toCH}$ is the average distance between member nodes in a cluster and the head node, it can be concluded that [10,21]:

$$d_{toBS} = 0.765\frac{M}{2} \tag{6}$$

$$d_{toCH} = \frac{M}{\sqrt{2\pi k}} \tag{7}$$

The number of optimal cluster heads is [13]:

$$k_{opt} = \frac{\sqrt{N}}{\sqrt{2\pi}}\sqrt{\frac{\varepsilon_{fs}}{\varepsilon_{mp}}\frac{M}{d_{toBS}^2}} \tag{8}$$

Therefore, the proportion of optimal cluster heads is:

$$p_{opt} = \frac{k_{opt}}{N} \tag{9}$$

In the initial stage of clustering, through broadcast among nodes in the networks, for any node $i$, there are a total of $n$ nodes within its communication range, of which the distance between $n_1$ nodes and $i$ is $< d_0$, and the distance between $n_2$ nodes and i is $> d_0$. So considering the ratio of the energy of $i$ and the average energy of all nodes within its communication range, $\varpi(E)_i$, the energy factor influencing the probability of cluster heads can be obtained:

$$\varpi(E)_i = \frac{E_i}{\sum_{j=1}^{n} E_j \Big/ n} \tag{10}$$

Consider the nodes distribution in the networks. If after nodes have been clustered, the average distance between nodes within the cluster and cluster head is far, a high communication cost is inevitable for one communication within the cluster. Set $\overline{E_{i-round}}$ is the average energy consumed in one communication between each node in the cluster and node $i$ after $i$ has been selected as the cluster head.

$$\overline{E_{i-round}} = \frac{\sum_{j=1}^{n_1}\left(l_j E_{elec} + l_j \varepsilon_{fs} d_{i,j}^2\right) + \sum_{k=1}^{n_2}\left(l_k E_{elec} + l_k \varepsilon_{mp} d_{i,k}^4\right)}{n} \tag{11}$$

On an ideal occasion that nodes in the networks are uniformly distributed and every data transmission send data identical in length $l$, the number of nodes in each cluster is $N/k_{opt}$. If the distance from $m_1$ nodes to the cluster head is $< d_0$, $m_2$ nodes $> d_0$, the ratio of these two types of nodes is :

$$\lambda = \frac{\pi d_0^2}{\pi d^2 - \pi d_0^2} \tag{12}$$

Therefore, the number of these two types of nodes is:

$$m_1 = \frac{\lambda \times \frac{N}{k_{opt}}}{\lambda + 1} \tag{13}$$

$$m_2 = \frac{\frac{N}{k_{opt}}}{\lambda + 1} \tag{14}$$

The random distribution of nodes can be viewed as a Poisson point process [21]. Ideally, there are $n$ points in circle $A$ and their locations which are uniformly distributed in $A$ are mutually independent random variables. $d_i$ is a random variable, presenting the distance from a point $(x_i, y_i)$ to the circle centre point. The expectation of all the points in the circle to the center point is:

$$E\left(d_i\right) = \iint_A \sqrt{x_i^2 + y_i^2}\,\rho(x_i, y_i)\mathrm{d}x\mathrm{d}y \tag{15}$$

A circle can be obtain after any radius revolves around the center, so consider the distribution of points on a random radius. Points are distributed uniformly in the circle, and accordingly, the density of points is proportional to radius squared. Therefore, the probability density of points on a random radius is:

$$f(x) = \frac{2x}{R^2} \tag{16}$$

Where R is radius length.

Therefore, the calculation of $E\left(d_i\right)$ can be simplified to:

$$E(d_i)=\int_0^R xf(x)\mathrm{d}x \tag{17}$$

By formula (16) and (17), the average distance expectation of nodes whose distance to the cluster head is less than $d_0$ is:

$$E(d_i)_1=\int_0^{d_0} x\frac{2x}{d_0^2}\mathrm{d}x=\frac{2}{3}d_0 \tag{18}$$

The average distance expectation of nodes whose distance to the cluster head is more than $d_0$ is:

$$E(d_i)_2=\int_{d_0}^{d} x\frac{2x}{(d-d_0)^2}\mathrm{d}x+d_0=\frac{2}{3}(d-d_0)+d_0 \tag{19}$$

Therefore, ideally the average energy consumption within one data transmission in the cluster is

$$\overline{E_{consume}}=\frac{l\left(m_1\left(E_{elec}+\varepsilon_{fs}E(d_i)_1\right)+m_2\left(E_{elec}+\varepsilon_{mp}E(d_i)_2\right)\right)}{\frac{N}{k_{opt}}} \tag{20}$$

By formula (11) and (20), communication cost factor $\varpi(C)_i$ which has influence on probability of cluster head election is:

$$\varpi(C)_i=\frac{\overline{E_{consume}}}{\overline{E_{i-round}}} \tag{21}$$

Integrating node energy factor and communication cost factor, the following formula can be used to calculate the probability for node i to become the cluster head node:

$$p_i=p_{opt}\times\left(\alpha\varpi(E)_i+\beta\varpi(C)_i\right) \tag{22}$$

Where $\alpha$ and $\beta$ are the calculation factors regulating the proportion of energy factor and communication cost factor in calculation $p_i$, $\alpha+\beta=1$.

The constraints of LEACH threshold formula $T(i)$ should be improved in two steps:

(1) to promote $T(i)$ into multi-level heterogeneous networks;

(2) in EEPCA, to take energy factor and the communication cost factors into account and to improve calculation method of $T(i)$, as is shown in formula (23):

$$T(i)=\begin{cases}\dfrac{p_i}{1-p_i\left(r \bmod \dfrac{1}{p_i}\right)}\left[\left(\alpha\varpi(E)_i+\beta\varpi(C)_i\right)+\left(r_s div\dfrac{1}{p_i}\right)\left(1-\left(\alpha\varpi(E)_i+\beta\varpi(C)_i\right)\right)\right], & \text{if } i\in G \\ 0, & \text{otherwise}\end{cases} \tag{23}$$

Where $r_s$ is the number of rounds when a node fails to be selected as the cluster head. Once the node elected, $r_s$ is reset to 0.

4.3 energy consumption prediction mechanism

Obviously, after the networks complete a round, a new node need to be selected as the cluster head. Because it is necessary to re-evaluate the energy factor and the communication cost factor so as to determine the probability for the node to become the cluster head, the current node residual energy must be obtained. The easiest way is that all nodes in the networks carry out a broadcast through the method utilized in the first round of clustering. However, considerable energy will be consumed when broadcasting in each round of clustering, so this paper establishes an energy consumption prediction mechanism for RDA nodes.

In $r-1$ round, it takes $n_j$ times for any node $j$ to send messages with a length $l_j$ to cluster head node $i$ and the distance between $i$ and $j$ is $d_{i,j}$. Since each node keeps relevant information of all nodes within communication range and their mutual distance, any node within node $j$ ' communication range can calculate the energy consumption of node $j$ in $r-1$ round.

$$E_{j_{r-1}_comsume}=\begin{cases} n_j\left(l_jE_{elec}+l_j\varepsilon_{fs}d_{i,j}^2\right), d_{i,j}<d_0 \\ n_j\left(l_jE_{elec}+l_j\varepsilon_{mp}d_{i,j}^4\right), d_{i,j}\geq d_0\end{cases} \tag{24}$$

According to the current energy of node $j$ and formula (24), the residual energy of node $j$ can be predicted at the beginning of $r$ round when $r-1$ round starts.

$$E_{j_r_prediction}=E_{j_{r-1}}-E_{j_{r-1}_comsume} \tag{25}$$

Due to reasons such as networks environment changes, when $r$ round starts, all nodes need to be re-clustered and new cluster head node need to be elected. Node $j$ determines whether its current residual energy is close to the residual energy predicted in the last round or not.

$$\gamma=\left|1-\frac{E_{j_r_prediction}}{E_{j_r}}\right| \tag{26}$$

If $\gamma$ is less than constant $\varepsilon$, the energy predication error can be tolerated. In the initial phase of $r$ round, node $j$ does not broadcast its energy information and the remaining nodes update node $j$ ' energy information in the routing table according to calculation results.

## 5. Simulation Experiment

5.1 establishment of simulation environment

Through simulation experiment, this paper makes analysis and comparison on the performance of EEPCA. The experiment simulates a high density sensor network for environmental monitoring randomly formed within a $100m \times 100m$ area. After the formation, nodes become static, no longer moving. And 100 sensor nodes are randomly distributed in this area, without loss of generality. Assuming the BS is located in the center of the area. In order to compare with other protocols, impact caused by random factors such as signal collision and wireless channel interference is ignored. Parameters used in this experiment can be seen in Table 1. This paper will compare the performance of EEPCA and that of LEACH, SEP and EDFCM. All results, unless otherwise stated, are average values of 100 times independent experiments.

Table 1 Parameters used in simulations

| Parameter | Value |
|---|---|
| Network grid | (0 ,0) ~(100 ,100) |
| node numbers | 100 |
| Coverage radius (m) | 12 |
| Threshold distance $d_0$(m) | 75 |
| Initial energy (J ) | 1-3 |
| $E_{elec}$ | 5nJ/bit |
| $\varepsilon_{fs}$ | 10pJ/bit/m$^2$ |
| $\varepsilon_{mp}$ | 0.0013pJ/bit/m$^4$ |
| Message size | 2000-6000bits |
| Broadcast Packet Sze | 2500 bits |
| Round | 5 TDMA frames |

5.2 Experiment Results and Analysis

In EEPCA, $\alpha$ and $\beta$ are calculation factors regulating the proportion of energy factor and communication cost factor in calculation $p_i$, satisfying $\alpha + \beta = 1$. Change the values of $\alpha$ and $\beta$ and observe the performance of EEPCA. This experiment sets all nodes are energy-heterogeneous and the initial energy is 1-3J. All monitored objects in the network are homogeneous, excluding RDA nodes. All nodes send 4000bits messages to the cluster head at TDMA timeslot.

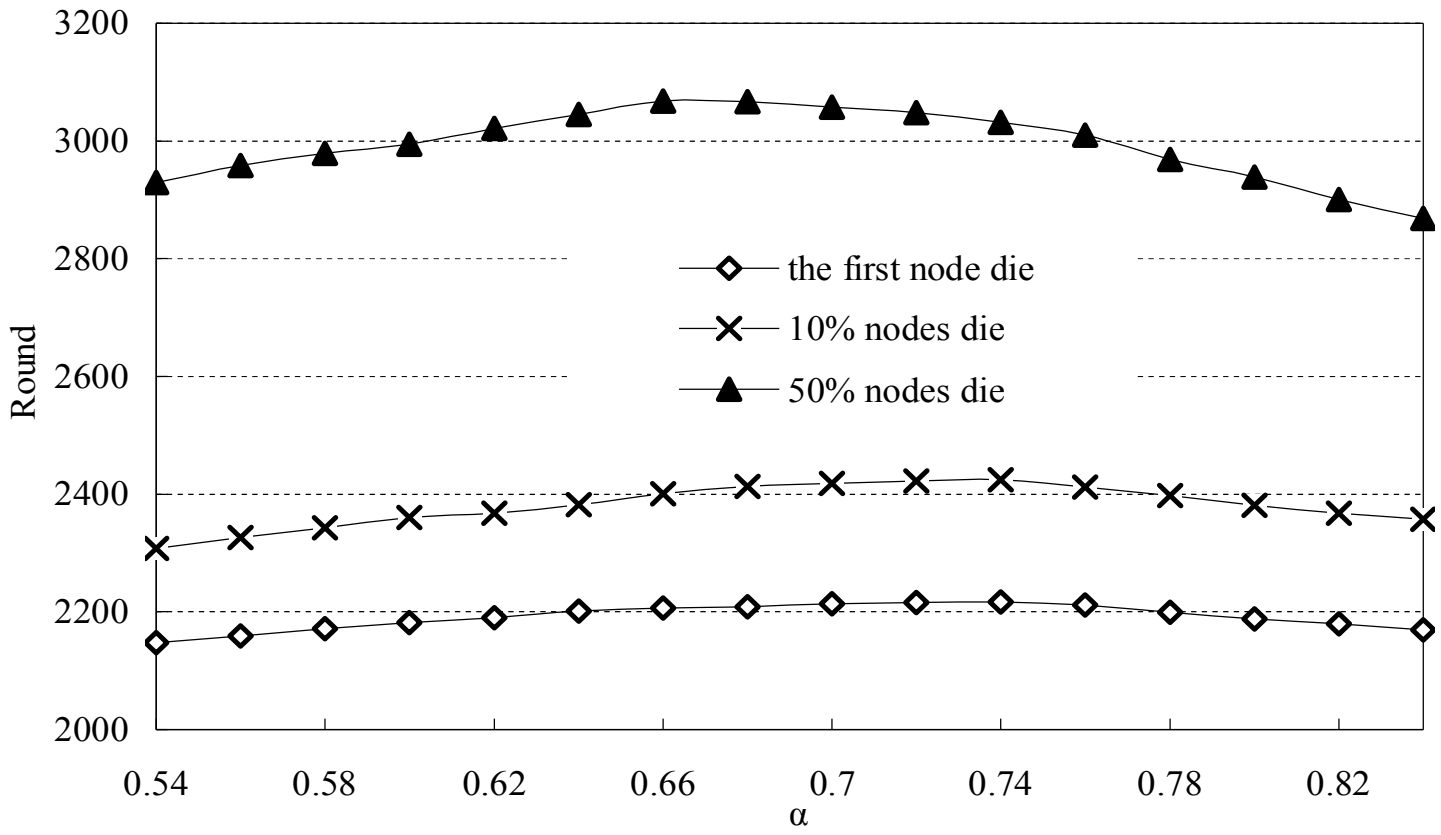

Fig.2 Influence of $\alpha$ and $\beta$ ' values on performance

Fig. 2 shows the death time of the first node, 10% nodes and 50% nodes when the values of $\alpha$ and $\beta$ vary in the above circumstances. It can be seen when the values of $\alpha$ are in the vicinity of 0.74, death time of the first node and 10% nodes appears the latest; while when the values of $\alpha$ are within the range of 0.66-0.68, death time of 50% nodes appears the latest. When parts of nodes in the network die, nodes density becomes significantly lower and due to the reduction of nodes number, network load is more likely to be uneven. Therefore, greater value of the communication cost factor $\beta$ can help improve algorithm performance. In subsequent experiments, the values of $\alpha$ and $\beta$ are unified as 0.7 and 0.3.

In the above experimental environment, EEPCA and LEACH, SEP and EDFCM will be compared and tested to analyze EEPCA cluster head selection mechanism's impact on the algorithm performance when all nodes are heterogeneous.

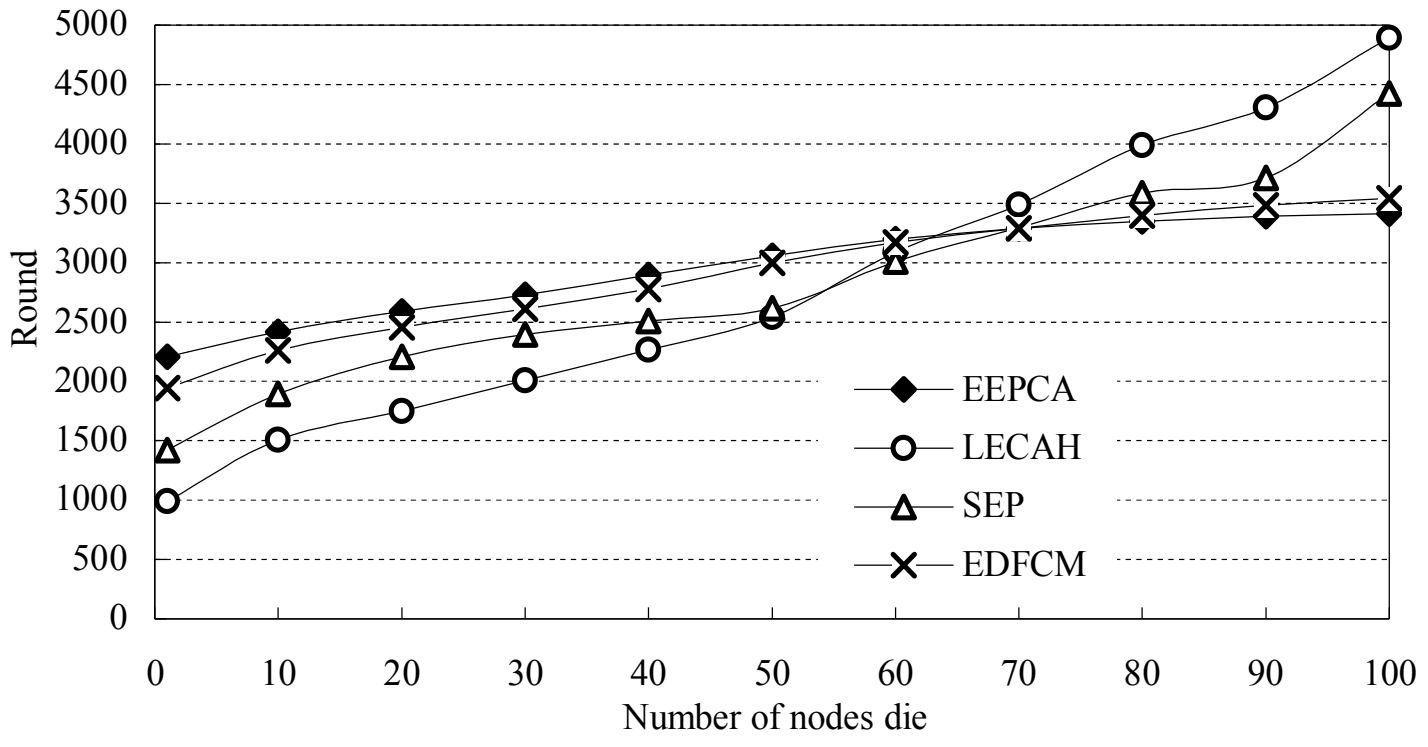

Fig.3 Death time of nodes

The simulation results in Fig. 3 show the variation of the number of dead nodes over time in the above experimental environment in different algorithms. It can be seen in Fig. 3, LEACH can not make good use of the additional energy of heterogeneous nodes, the stable period is very short and nodes die at a fixed speed rate. Compared with LEACH, SEP has longer stable periods. EEPCA and EDFCM curves are lines with smaller slope versus X-axis. Because EEPCA distributes energy consumption uniformly on each node in the heterogeneous network, the death time of the first and the last node is relatively closer. It can be seen from Fig. 3, compared with LEACH and SEP, EEPCA can prolong network life expectancy by 129% and 55%.

In the above experimental environment, change the proportion of heterogeneous nodes in the total number of nodes and observe the performance of each algorithm. Fig. 4 presents the number of rounds from the beginning to the death of the first node when the proportion of heterogeneous nodes varies from 0 to 100%. In this experiment, the initial energy of all non-energy-heterogeneous nodes is 2J.

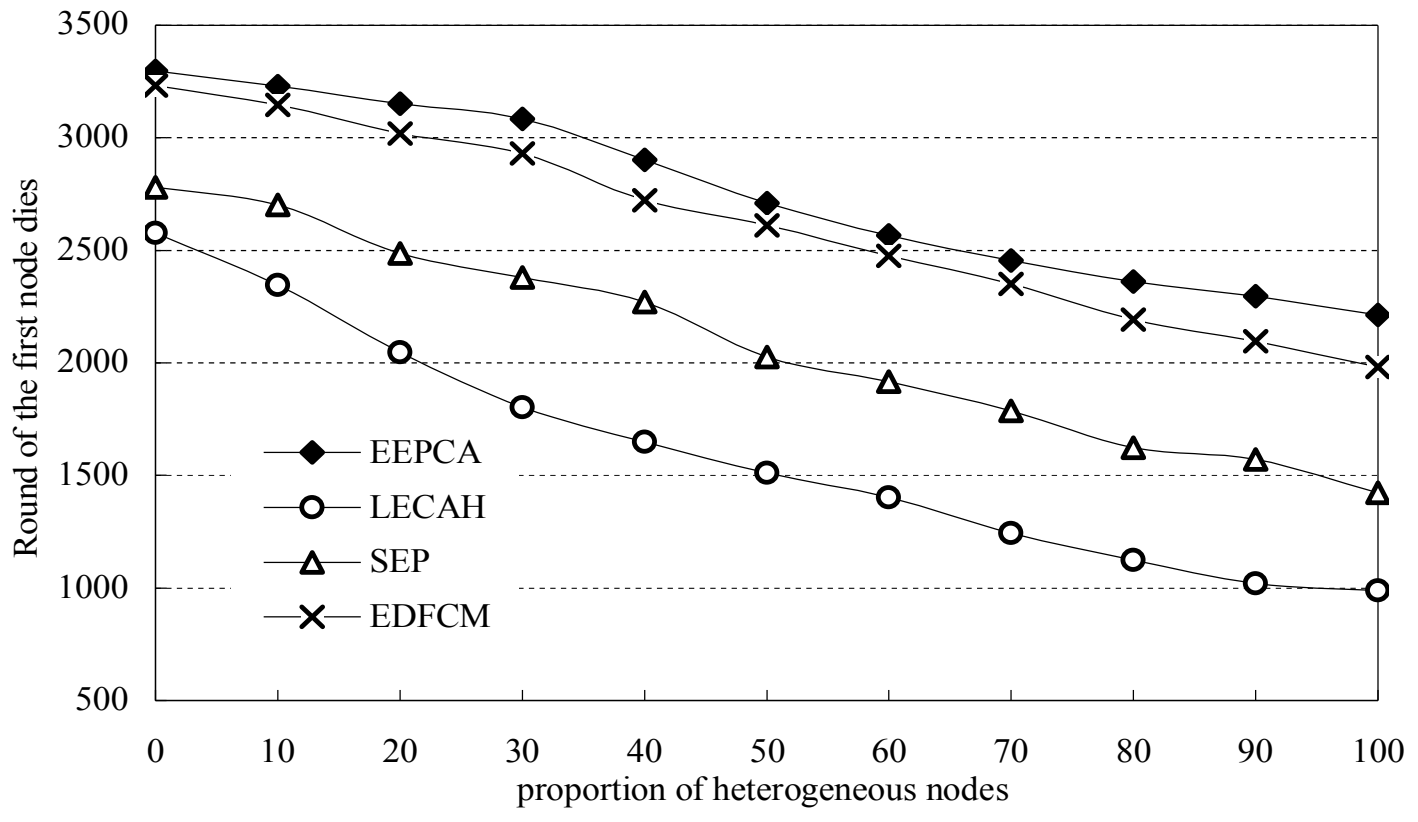

Fig.4 Death time of the first node when the number of heterogeneous nodes changes

Before the death of 10% nodes, the network can send back to the BS data of high quality and reliability [13]. So Fig. 5 presents the number of rounds from the beginning to the death of 10% nodes, namely the stable period.

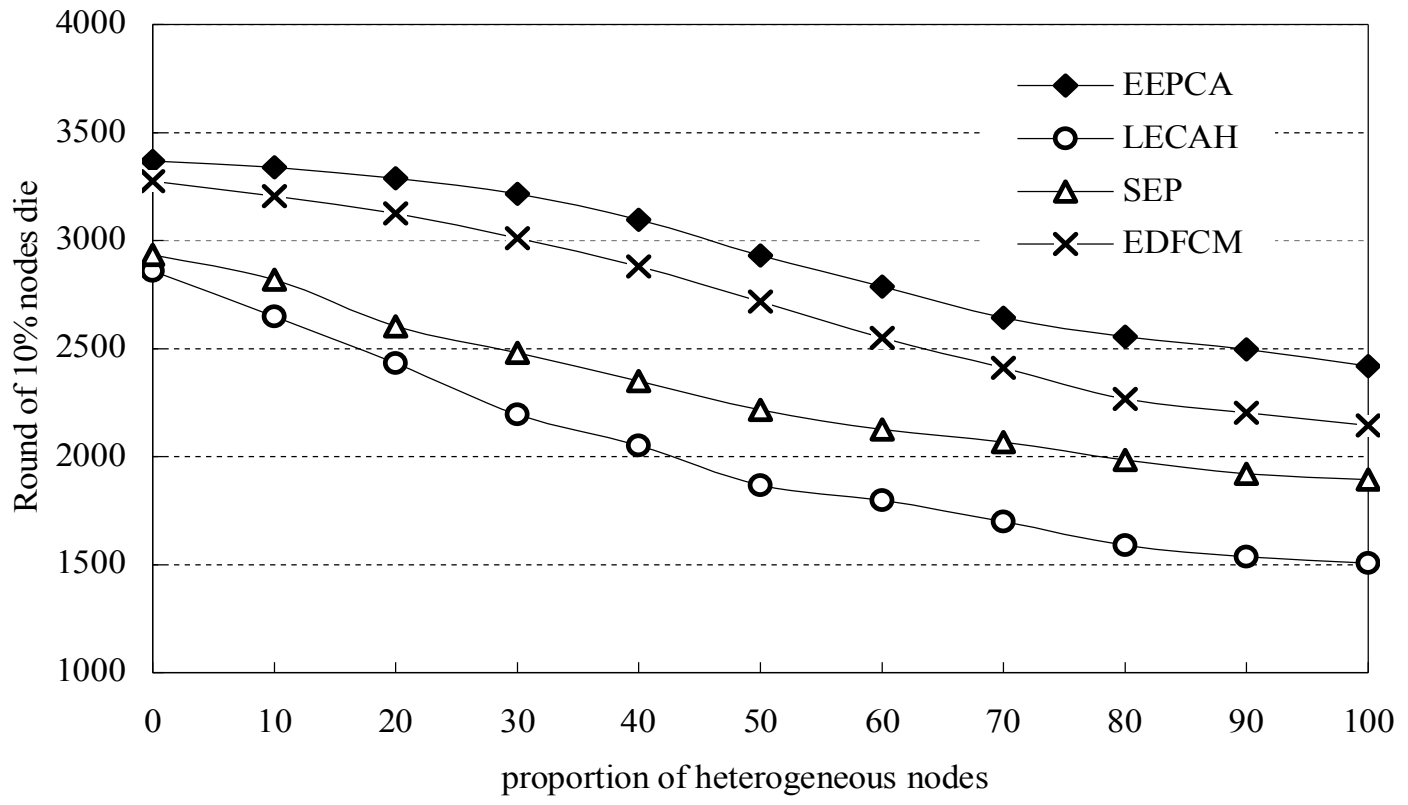

Fig.5 network stable period when the number of energy heterogeneous nodes changes

It can be seen that as LEACH is not a clustering algorithm for heterogeneous networks, it does not take into account the energy difference between nodes and instead, all nodes are treated equally. Therefore, in LEACH, with the increase of the proportion of heterogeneous nodes, attainable network stable period quickly reduces. SEP can obtain 25% more stable period than LEACH, which is basically consistent with the experimental results presented by [11]. As EDFCM takes into account heterogeneous energy of different nodes, the death time of its first node is later than SEP and it gets longer stable period than SEP. EEPCA takes into account the energy consumption of nodes in the communication process in addition to residual energy, so the decline rate of stable period is significantly less than other algorithms in the process of increasing proportion of heterogeneous nodes. Therefore, with greater proportion of heterogeneous nodes, a more stable period is obtained.

To go further, RDA nodes are introduced into the experiment. Set all nodes energy in the networks is heterogeneous and 50% nodes are RDA nodes. Meanwhile, because of factors such as changes in the environment, 10% nodes are malfunctioning. All RDA nodes send messages 3-7 times in a round and the sizes of messages are valued randomly between 2000-6000bits. Examine the impact of the constant $\varepsilon$ on networks stable period.

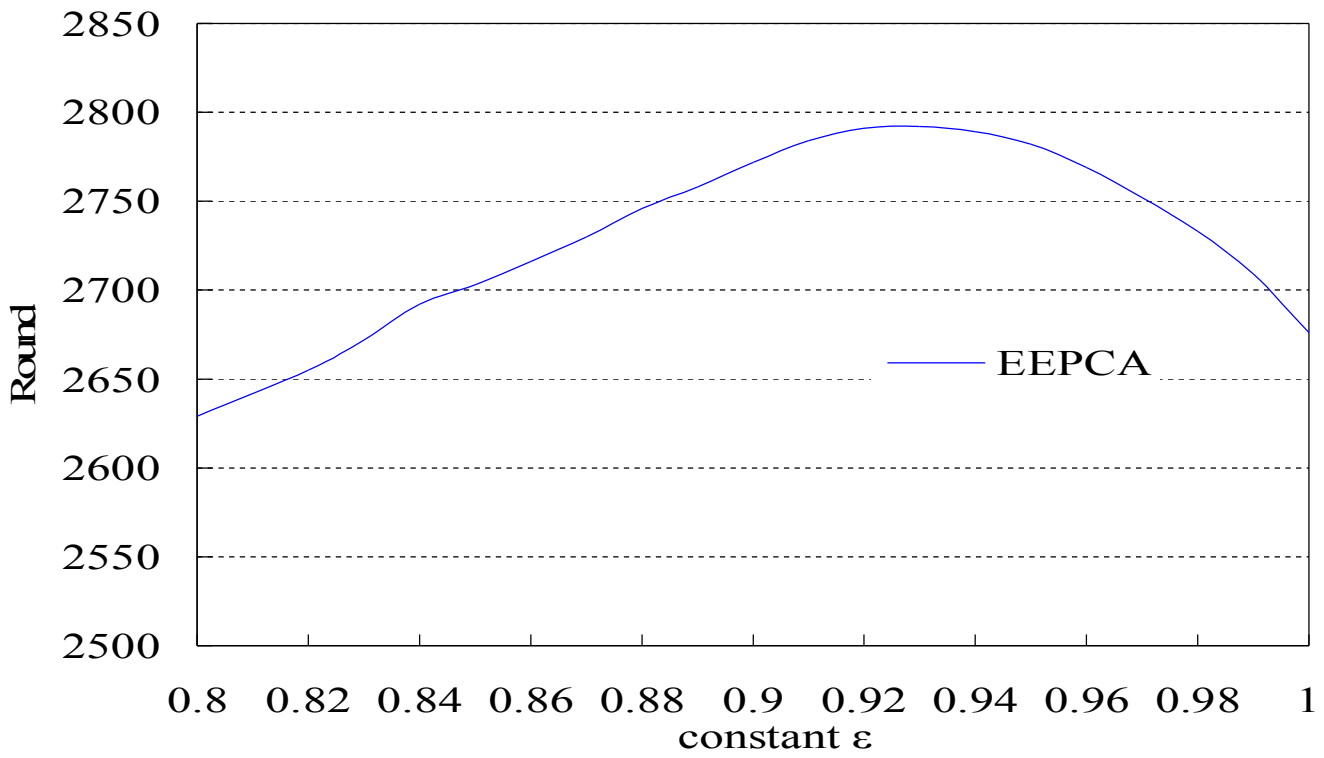

Fig.6 Impact of constant $\varepsilon$ on network stable period

Due to malfunctioning nodes in the network, errors in energy prediction are inevitable. If $\varepsilon = 1$, nodes broadcast their energy information when energy prediction errors happen. In this case, it is difficult to achieve substantial savings in energy consumption. If the value of $\varepsilon$ is too low, nodes do not broadcast their energy information even if biggish errors in actual residual energy and predicted energy happen. In this case, nodes with lower actual residual energy may have higher opportunity to become the cluster head and the length of network stable period is thus affected. Fig. 4 shows when the value of $\varepsilon$ is near 0.92-0.93, the network achieves maximum stable period.

RDA nodes are introduced into LEACH, SEP, EDFCM and EEPCA and the stable periods of all these algorithms are examined. This experiment sets all nodes are energy heterogeneous, 50% of which are RDA nodes, constant $\varepsilon = 0.93$ and 10% nodes in the network are malfunctioning. The results are shown in Fig. 7:

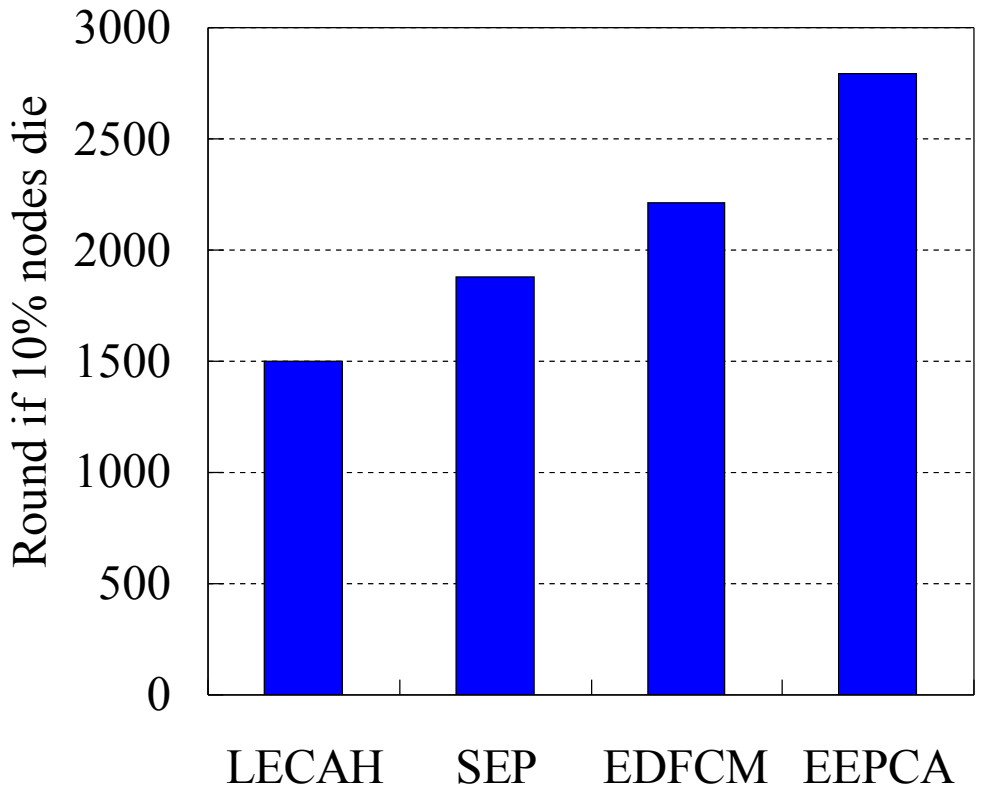

Fig.7 network stable period

Obviously, due to the introduction of energy consumption prediction mechanism, broadcast frequency in the clustering phase in each round is effectively reduced. Therefore, in a network heterogeneous in two ways --- initial energies and monitored objects, EEPCA makes significant improvement in network stable period compared with the other three algorithms. However, the heterogeneity of EDFCM fails to take into account RDA nodes, so when these nodes are added, the stable period of EDFCM declines considerably.

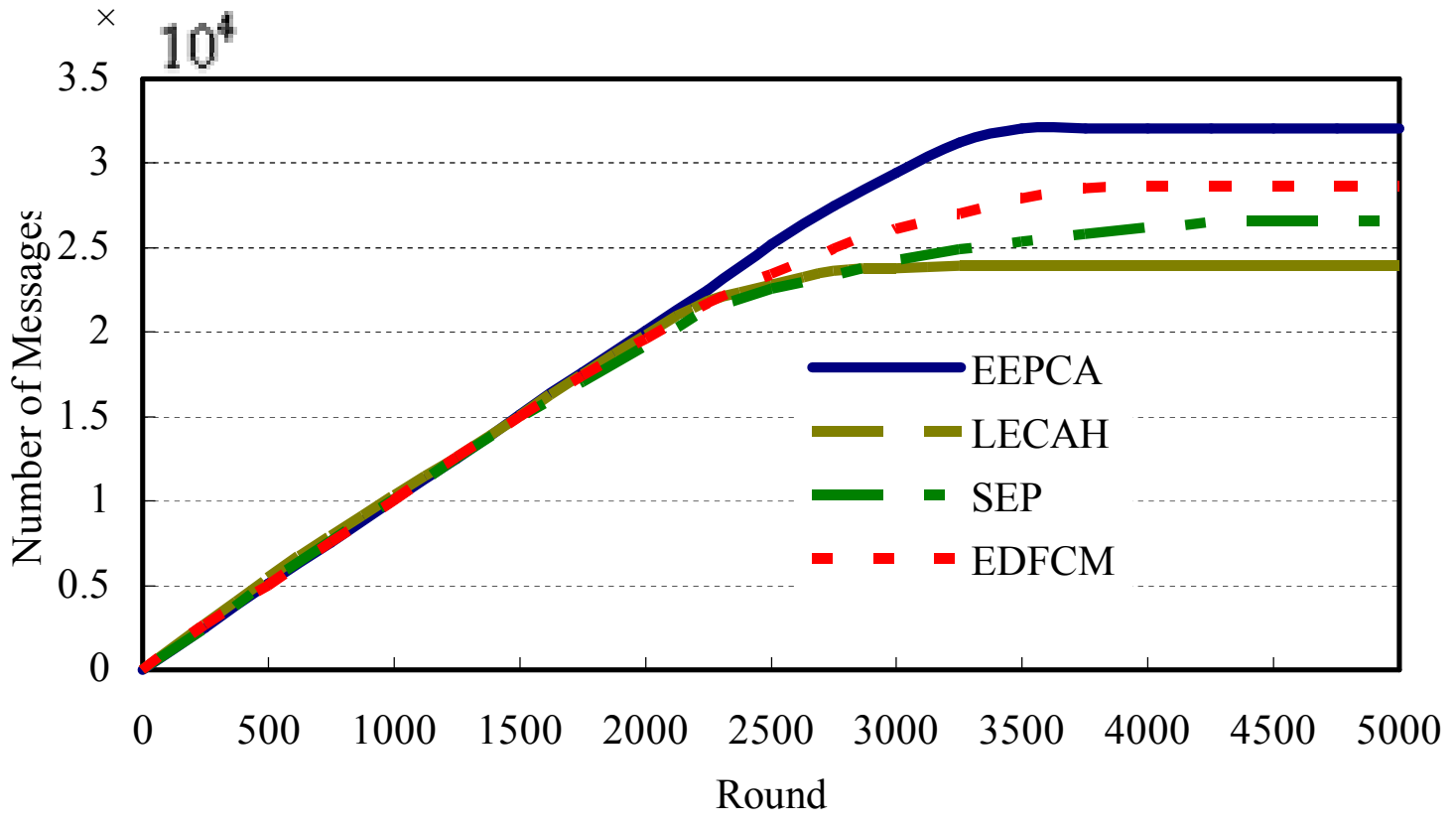

Fig.8 the number of messages received by BS

For the algorithm running by round, monitoring quality can be measured by the total times for all nodes in the network to collect data. Fig. 8 shows that all the nodes are energy heterogeneous, 50% are RDA nodes and 10% of the nodes are malfunctioning. In EEPCA, the number of messages received by BS is on linear rise for a long period of time, while in other algorithms, the growth rate of the number of messages received by BS begins to decline earlier. To sum up the total number of messages sent back to BS by all nodes in these four algorithms when the network fails, the amount of data collected by EEPCA is much larger than that by the other three algorithms. Therefore, EEPCA has better network monitoring quality.

## 6 Conclusion

In this paper, we describe the HWSN model with both different initial energies and monitored objects. We present an effective energy prediction clustering algorithm EEPCA for multi-level heterogeneous sensor networks. In EEPCA, each node independently selects itself as the cluster head node based on energy factor and communication cost factor, which leads to the probability of cluster head election related to nodes' current residual energy and average communication cost after being selected. At the same time, with the consideration that the WSN are frequently used to monitor objects such as temperature and humidity which need to report data regularly, and the length of reported data are usually fixed, an energy consumption prediction mechanism is established for RDA nodes. Simulation results show that compared with LEACH, SEP and EDFCM, EEPCA can achieve longer lifetime, higher energy efficiency and better network monitoring quality. Its performance is superior to other protocols.

In future work, research will further improve residual energy prediction mechanisms so as to achieve greater prediction accuracy and prolong network lifetime to the maximum. In addition, such problems will be considered as message transmission and energy prediction in networks where one node monitors a variety of different objects. Our ultimate goal is to apply EEPCA algorithm to practical use.

## Acknowledgements

This work is supported by the National Natural Science Foundation of China (61003310), the Opening Project of State key Laboratory of Networking and Switching Technology (Beijing University of Posts and Telecommunications)(SKLNST-2010-1-03), the Scientific Research Fund of SiChuan Provincial Education Department(10ZB005).

## References

[1] I.F. Akyildiz, W. Su, Y.Sankarasubramaniam, E.Cayirici, Wireless sensor network: a survey. Computer Networks, 38(4) (2002) 393-422.

[2] M. Haenggi, Handbook of Sensor Networks: Compact Wireless and Wired Sensing Systems, CRC Press, 2005.pp.1-11.

[3] C.Y. Chong, S.P. Kumar, Sensor networks: evolution opportunities, and challenges, Proceedings of IEEE 91 (8) (2003) 1247–1256.

[4] D. Estrin, L. Girod, G. Pottie, M. Srivastava, Instrumenting the world with wire- less sensor networks, in: In International Conference on Acoustics, Speech, and Signal Processing (ICASSP 2001). (2001) 2033–2036.

[5] C.Y. Chang, H.R. Chang, Energy-aware node placement, topology control and MAC scheduling for wireless sensor networks. Computer Networks. 52(11)(2008) 2189–2204.

[6] E.J. Duarte-Melo, M.-Y. Liu, Analysis of energy consumption and lifetime of heterogeneous wireless sensor networks, in Proc. of the GLOBECOM, New York: IEEE Press.(2002) 21-25.

[7] E.P.de Freitas, T. Heimfarth, C.E.Pereira, Evaluation of coordination strategies for heterogeneous sensor networks aiming at surveillance applications, in:Proceedings of IEEE Sensors (SENSORS), Christchurch, New Zealand.(2009) 591–596.

[8] J.M.Corchado, J.Bajo, D.I.Tapia, A.Abraham, Using heterogeneous wireless sensor networks in a telemonitoring system for healthcare, IEEE Transactions on Information Technology in Biomedicine 14 (2) (2010) 234–240.

[9] I.Dietrich, F.Dressler, On the lifetime of wireless sensor networks, ACM Transactions on Sensor Networks 5 (1) (2009) 5:1–5:39.

[10] W.R.Hernzelman, A.P.Chandrakasan, H.Balakrishnan, An application specific protocol architecture for wireless microsensor networks. IEEE Transactions on Wireless Communications, 1 (4) (2002) 660-670.

[11] G. Smaragdakis, I. Matta, A. Bestavros, SEP: a stable election protocol for clustered heterogeneous wireless sensor networks, in: Proceedings of the International Workshop on SANPA, 2004. pp. 251–261.

[12] V.P. Mhatre, C. Rosenberg, D. Kofman, A minimum cost heterogeneous sensor network with a lifetime constraint, IEEE Transactions on Mobile Computing 4 (1) (2005) 4–14.

[13] Q. Li, Q.X. Zhu, M.W. Wang, Design of a distributed energy-efficient clustering algorithm for heterogeneous wireless sensor networks, Computer Communications 29 (12) (2006) 2230–2237.

[14] K. Dilip, C.A. Trilok, R.B. Patel, EEHC: energy efficient heterogeneous clustered scheme for wireless sensor networks, Computer Communications 32 (4) (2009) 662–667

[15] J.M. Kim, S.H. Park, Y.J. Han, T.M. Chung, CHEF: cluster head election mechanism using fuzzy logic in wireless sensor networks, in: Proceedings of the ICACT, February 2008, pp. 654–659.

[16] K. Dilip, C.A. Trilok, R.B. Patel, EEHC: energy efficient heterogeneous clustered scheme for wireless sensor networks, Computer Communications 32 (4) (2009) 662–667

[17] H.B. Zhou, Y.M, Y.Q Hu, G.Z. Xie, A novel stable selection and reliable transmission protocol for clustered heterogeneous wireless sensor networks. Computer Communications. 33(2010):1843-1849.

[18] B.A. Attea, E.A. Khalil, A new evolutionary based routing protocol for clustered heterogeneous wireless sensor networks. Applied Soft Computing.(2011).

[19] S. Jin, M. Zhou, A.S. Wu, Sensor network optimization using a genetic algorithm, in: Proceedings of the 7th World Multiconference on Systemics, Cybernetics and Informatics, 2003.

[20] S.Doshi, S. Bhandare, T.Brownl, An on-demand minimum energy routing protocol for a wireless ad hoc network. ACMSIGMOBIL E Mobile Computing and Communications Review, 6(3) (2002) 50-66.

[21] V.Mhatre, C.Rosenberg, Design guidelines for wireless sensor networks: communication, clustering and aggregation. Ad Hoc Network Journal, 2(1) (2004) 45−63.